\documentclass[a4paper,aps,pra,preprintnumbers]{revtex4}
\usepackage[utf8]{inputenc}
\usepackage[english]{babel}
\usepackage[T1]{fontenc}
\usepackage{amssymb,amsfonts,amsmath,mathtext,enumerate,float,dsfont}
\usepackage{graphics,graphicx,epsfig,epstopdf}
\usepackage{caption}
\usepackage{cmap}
\usepackage{multirow}
\usepackage{indentfirst}
\usepackage[usenames]{color}
\usepackage{amsthm}
\usepackage{xcolor}

\begin{document}
\title{Bidirectional quantum teleportation in continuous variables}
\author{E. A. Nesterova$^{1}$, S. B. Korolev$^{1,2}$} 
\affiliation{$^1$ St.Petersburg State University, Universitetskaya nab. 7/9, St.Petersburg, 199034, Russia}
\affiliation{$^2$ Laboratory of Quantum Engineering of Light, South Ural State University, pr. Lenina 76, Chelyabinsk, 454080, Russia}
\date{\today}

\begin{abstract}
    We propose a bidirectional quantum teleportation protocol in continuous variables. We use a cluster state in continuous variables as the main resource to realize this protocol. In the paper, we obtain a family of configurations of cluster states in continuous variables that can be used to realize the bidirectional quantum teleportation protocol. From the whole family of configurations, we have chosen those that realize the protocol with the smallest possible error.
\end{abstract}
\maketitle
\section{Introduction}
The quantum teleportation protocol was originally designed to transfer an unknown quantum state from one user (Alice) to another (Bob) \cite{Bennett}. However, this protocol has the limitation that teleportation is only possible in one direction, from Alice to Bob. To exchange quantum information simultaneously, Alice and Bob need to use a different protocol called bidirectional quantum teleportation (BQT). Currently, the BQT protocol is actively being developed \cite{Zha_BDQT,Fu2014,Li_GHZ,Yan2013,Duan2014,Chen2014}.  

In works \cite{Zha_BDQT,Fu2014,Li_GHZ,Yan2013,Duan2014,Chen2014} devoted to BQT, discrete-variable physical systems described have been proposed to operate the protocol. However, the main limitation of such systems is their probabilistic entanglement with each other. Since BQT requires multipartite entangled states \cite{Fu2014}, generating the main resource of BQT in discrete variables becomes a problem. 

To solve this problem, one can use physical systems in continuous variables that are deterministically entangled \cite{Braunstein_RevModPhys.77.513}. In \cite{Braunstein_telep_cv}, it was shown that it is possible to implement traditional teleportation of quantum states in continuous variables. All this means that physical systems in continuous variables can potentially be used for deterministic BQT.

However, continuous-variable physical systems also have drawbacks. To realize teleportation, the physical systems involved need to be in squeezed states that are entangled with each other, forming the main resource, which is a multi-particle entangled state. When using such a resource, the teleported state will have an error that is proportional to the degree of squeezing of the states used, as well as to the number of physical systems involved \cite{Korolev_2020_1}.

The most obvious way to reduce the teleportation error in continuous variables is to use states with a higher squeezing degree. However, it is very problematic to create high squeezing states experimentally \cite{Vahlbruch_2016}. Another way to reduce the error is to optimize the resource multipartite entangled state \cite{Korolev_2020}. The process of such optimization includes finding the optimal number of squeezed states, which on the one hand are necessary for the implementation of BQT, and on the other hand, give the smallest error. In addition, it is necessary to choose the correct configuration (entanglements between physical systems) of the resource state.

In our work, we will propose a BQT scheme in continuous variables. We will analyze the teleportation errors obtained in our scheme and optimize the teleportation protocol so that the error is minimally possible.

\section{The main resource for bidirectional quantum teleportation in continuous variables}

The goal of our work is to generalize the traditional continuous variable teleportation protocol so that both Alice and Bob can simultaneously transfer unknown states to each other.

As in traditional teleportation \cite{Bennett,Braunstein_telep_cv}, the main resource for generalized teleportation must be an entangled state. Since in the generalized teleportation protocol, both Alice and Bob are simultaneously senders and receivers, they need a multipartite entangled state as a resource. We will use a cluster state as this resource. 

A cluster state is a quantum multipartite entangled state characterized by a mathematical graph. The nodes of the graph are physical systems in continuous variables, while the edges represent entanglements between them. A special feature of cluster states in continuous variables is that the graph is weighted \cite{Korolev_2018,Zinatullin_2022}. This means that each edge of the graph is associated with a real number, called the weight coefficient.

The physical systems used to generate a cluster state are quantum oscillators in quadrature-squeezed states. Each such state is described by two observable operators $\hat{x}$ and $\hat{y}$. These operators obey the canonical commutation relation:
 \begin{align}
\left[ \hat{x}_j,\hat{y}_k\right]=\frac{i}{2}\delta _{j,k},
\end{align}
where the indices $j$ and $k$ enumerate the corresponding oscillators and $\delta _{j,k}$ represents the Kronecker delta. We assume that all oscillators are squeezed in the $\hat y$-quadrature \cite{Gu}. This means that their variances are smaller than the variances of the vacuum state:
\begin{align} 
\langle \delta \hat{y}_j^2 \rangle < \frac{1}{4}, \qquad j=1,\dots,n.
\end{align}

To obtain the cluster state, characterized by the graph G, we have to apply a Bogoliubov transformation \cite{Bogoljubov} to independent quantum oscillators. This transformation is given by the matrix \cite{Korolev_2018,Ukai_1}:
\begin{align} \label{U_Bogol}
U=(I+iA)(I+A^2),
\end{align}
where $A$ is the adjacency matrix of the graph corresponding to the generated cluster state. The adjacency matrix completely determines the cluster state graph. The elements of this matrix are the weight coefficients of the cluster state graph $g_{ij}$. If the $i$-th and $j$-th nodes of the graph are not connected by an edge, then $g_{ij}=0$.

\section{Bidirectional Quantum Teleportation Protocol on a Five-Node Cluster State in Continuous Variables}
\subsection{The main result of teleportation}
Let us now proceed directly to the implementation of the BQT protocol. As the main resource for realizing such teleportation, let us first consider a cluster state containing five nodes in the graph. A cluster state with such a number of nodes was considered in the paper on BQT in discrete variables \cite{Zha_BDQT}. However, we will not specify the graph of this state for now. We will consider the protocol on a cluster state with a complete weighted graph and further optimize the weight coefficients to obtain bidirectional teleportation between Alice and Bob.

\begin{figure}[H]
    \centering
    \includegraphics{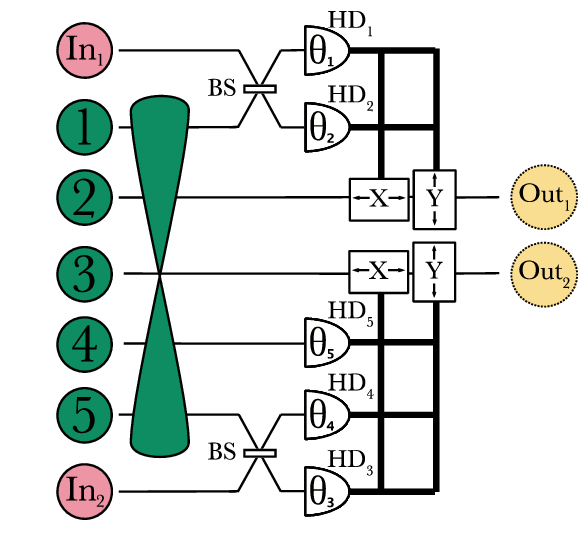}
    \caption{Scheme of bidirectional quantum teleportation. In the figure, modes 1-5 are entangled modes of the cluster state; $\text{In}_i$ are input teleported states; BS is a symmetric beam splitter; $\text{HD}_i$ is a homodyne detector whose local oscillator has a phase of $\theta_i$; $\text{Out}_i$ are output states.}
    \label{fig:bdt}
\end{figure}

The entire BQT protocol in continuous variables is depicted in Fig. \ref{fig:bdt}.
In the protocol, the five-node cluster state is distributed between Alice and Bob. Alice holds the states in nodes 1 and 3, while Bob holds the states in nodes 5 and 2. Node 4 is an auxiliary. For the scheme's symmetry, the state from this node is given to Charlie, who acts as a helper in this protocol. It should be noted that in the real scheme, the state of this node can be given to either Alice or Bob, which will not affect the result.

Alice and Bob entangle the quantum states they want to teleport with states in nodes 1 and 5, respectively. For this purpose, they use a symmetric beam splitter. After that, the process of measurements takes place using homodyne detectors. All states are measured except for the states at nodes 2 and 3. The last step of the teleportation is the displacement of the quadratures of the unmeasured states. The displacement occurs according to the results of measurements obtained by Alice, Bob, and Charlie. Let us now describe this teleportation procedure in more detail.

\begin{figure}[H]
    \centering
    \includegraphics[scale=0.3]{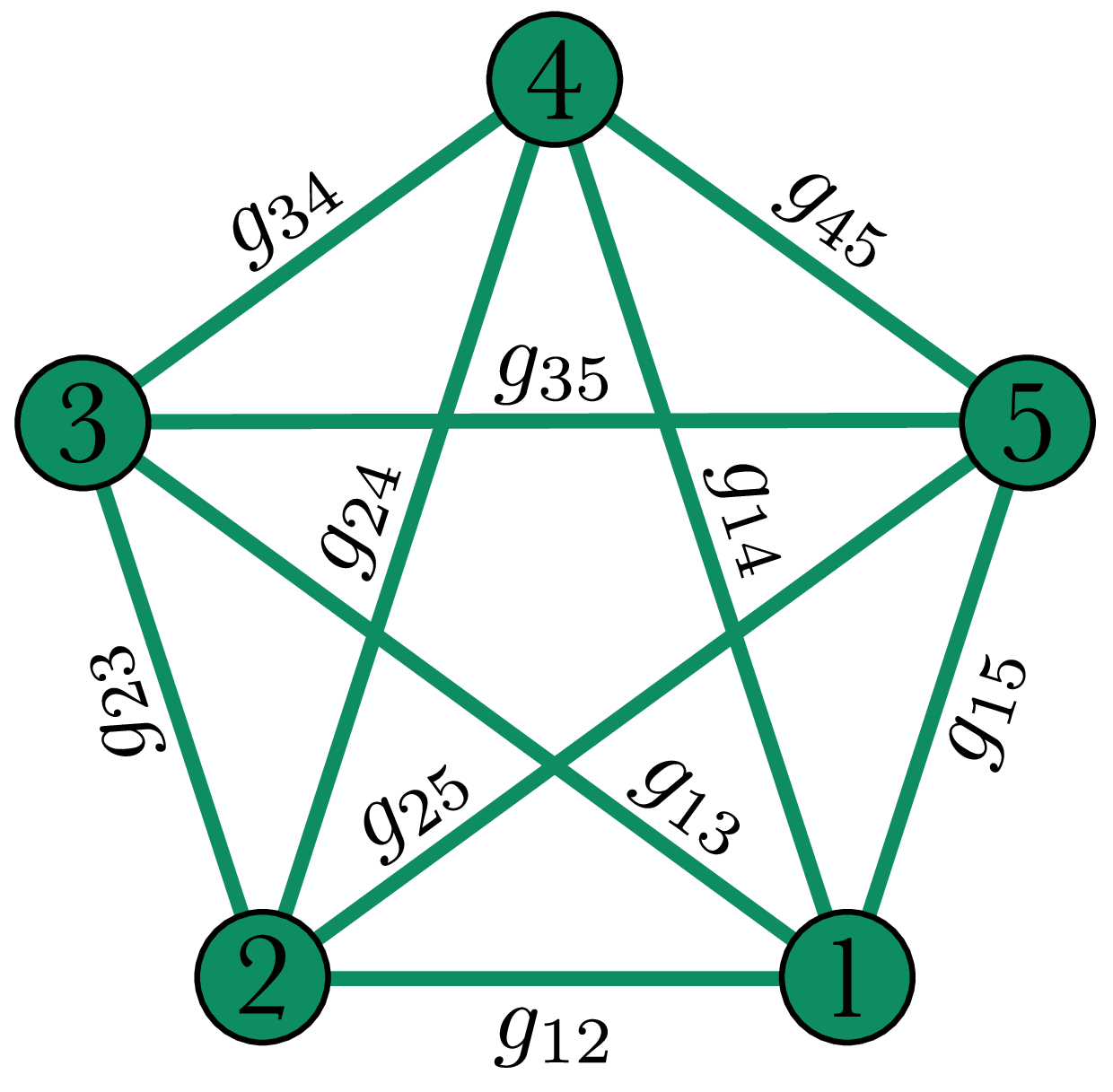}
    \caption{Five-node complete weighted graph of cluster state.}
    \label{fig:graph}
\end{figure}

First, we need to generate a five-node cluster state. As mentioned earlier, we first consider a cluster state whose graph is a complete weighted graph. The graph of such a cluster state is shown in Fig. \ref{fig:graph}.
The adjacency matrix of such a graph is given as follows:
\begin{equation}
A_5 = \left(
\begin{array}{ccccc}
0 & g_{12}& g_{13}& g_{14}& g_{15}\\
g_{12} & 0& g_{23}& g_{24}& g_{25}\\
g_{13} & g_{23}& 0& g_{34}& g_{35}\\
g_{14} & g_{24}& g_{34}& 0& g_{45}\\
g_{15} & g_{25}& g_{35}& g_{45}& 0\\
\end{array}
\right).
\end{equation}
Knowing the adjacency matrix, one can calculate the matrix of the Bogoliubov transformation (\ref{U_Bogol}), which is necessary to perform over independent squeezed quantum oscillators to create the cluster state. The relation between the quadrature operators of the cluster state and the quadrature operators of the independent squeezed quantum oscillators is given by the following relation:
\begin{align}
\hat{\vec{X}}+i\hat{\vec{Y}}=\left(I+iA_5\right)\left(I+A_5^2\right)\left( \hat{\vec{x}}_s+i\hat{\vec{y}}_s\right),
\end{align}
where $\hat{\vec{X}}=\left(\hat{X}_1,\hat{X}_2,\dots, \hat{X}_5\right)^T$ and $\hat{\vec{Y}}=\left(\hat{Y}_1,\hat{Y}_2,\dots, \hat{Y}_5\right)^T$ are vectors consisting of quadratures of the cluster state; $\hat{\vec{x}}_s=\left(\hat{x}_{s,1},\hat{x}_{s,2},\dots, \hat{x}_{s,5}\right)^T$ and $\hat{\vec{y}}_s=\left(\hat{y}_{s,1},\hat{Y}_{s,2},\dots, \hat{y}_{s,5}\right)^T$ are vectors consisting of quadratures of the squeezed quantum oscillators. To simplify the notation, we introduce new quadratures that are related to the quadratures of squeezed oscillators as follows: $\hat{\vec{x}}_r+i\hat{\vec{y}}_r=\left(I+A_5^2\right)\left( \hat{\vec{x}}_s+i\hat{\vec{y}}_s\right)$. Taking into account these reductions, the vector of quadratures of the cluster state can be written as follows:
\begin{equation} \label{eq_0}
\begin{pmatrix}
\hat{X}_1+i\hat{Y}_1\\
\hat{X}_2+i\hat{Y}_2\\
\hat{X}_3+i\hat{Y}_3\\
\hat{X}_4+i\hat{Y}_4\\
\hat{X}_5+i\hat{Y}_5\\
\end{pmatrix}=
\begin{pmatrix}
\hat{x}_{r,1}+i\hat{y}_{r,1} + g_{12}(i\hat{x}_{r,2}-\hat{y}_{r,2}) + g_{13} (i\hat{x}_{r,3}-\hat{y}_{r,3}) + g_{14} (i\hat{x}_{r,4}-\hat{y}_{r,4}) + g_{15} (i\hat{x}_{r,5}-\hat{y}_{r,5})\\
\hat{x}_{r,2}+i\hat{y}_{r,2} + g_{12}(i\hat{x}_{r,1}-\hat{y}_{r,1}) + g_{23} (i\hat{x}_{r,3}-\hat{y}_{r,3}) + g_{24} (i\hat{x}_{r,4}-\hat{y}_{r,4}) + g_{25} (i\hat{x}_{r,5}-\hat{y}_{r,5})\\
\hat{x}_{r,3}+i\hat{y}_{r,3} + g_{13}(i\hat{x}_{r,1}-\hat{y}_{r,1}) + g_{23} (i\hat{x}_{r,2}-\hat{y}_{r,2}) + g_{34} (i\hat{x}_{r,4}-\hat{y}_{r,4}) + g_{35} (i\hat{x}_{r,5}-\hat{y}_{r,5})\\
\hat{x}_{r,4}+i\hat{y}_{r,4} + g_{14}(i\hat{x}_{r,1}-\hat{y}_{r,1}) + g_{24} (i\hat{x}_{r,2}-\hat{y}_{r,2}) + g_{34} (i\hat{x}_{r,3}-\hat{y}_{r,3}) + g_{45} (i\hat{x}_{r,5}-\hat{y}_{r,5})\\
\hat{x}_{r,5}+i\hat{y}_{r,5} + g_{15}(i\hat{x}_{r,1}-\hat{y}_{r,1}) + g_{25} (i\hat{x}_{r,2}-\hat{y}_{r,2}) + g_{35} (i\hat{x}_{r,3}-\hat{y}_{r,3}) + g_{45} (i\hat{x}_{r,4}-\hat{y}_{r,4})\\
\end{pmatrix}.
\end{equation}

\begin{figure}[H]
    \centering
    \includegraphics[scale=0.3]{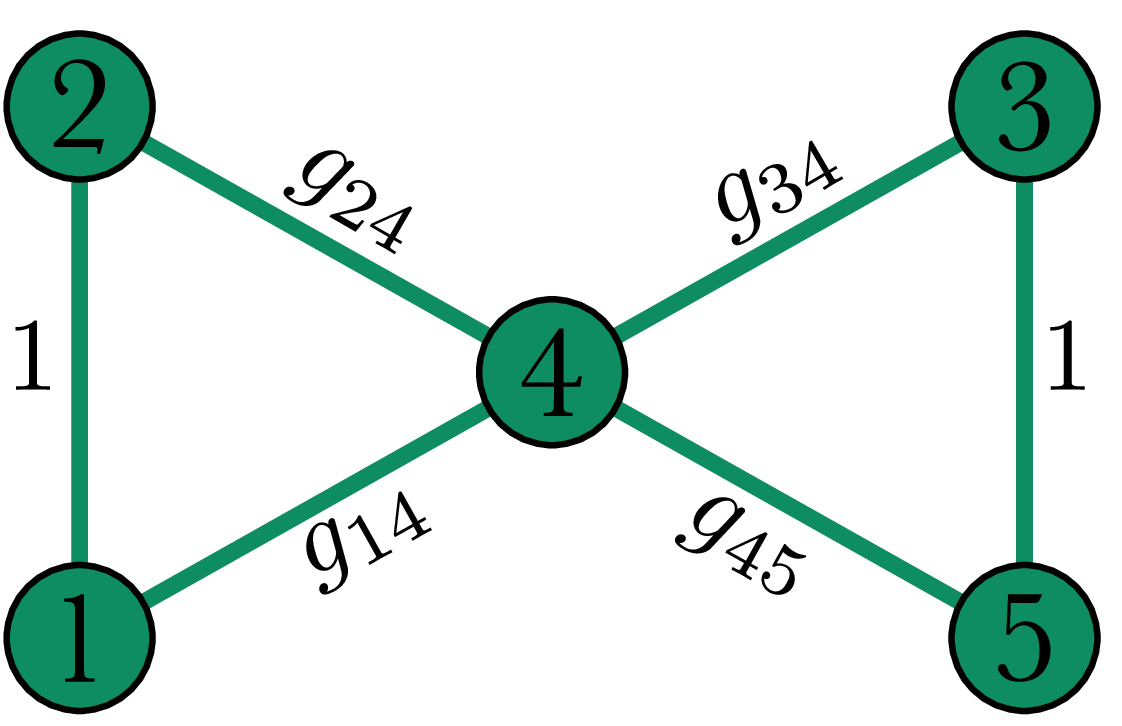}
    \caption{Five-node cluster state suitable for implementing bidirectional quantum teleportation in continuous variables.}
    \label{fig:TWT}
\end{figure}

Once we have obtained the cluster state, we can proceed with the teleportation procedure. For teleportation, Alice and Bob take two quantum states. The teleported state of Alice is described by the annihilation operator $\hat{a}_{in}=\hat{x}_a+i\hat{y}_a$, and the teleported state of Bob is $\hat{b}_{in}=\hat{x}_b+i\hat{y}_b$. To entangle their teleported states, Alice and Bob use symmetric beam splitters whose transformation matrix has the form: 
\begin{align}
    U_{BS}=\frac{1}{\sqrt{2}}\begin{pmatrix}
        1 & 1\\
        1 & -1
    \end{pmatrix}.
\end{align}
Alice entangles her teleported state with the state in the first node of the cluster state, and Bob Alice entangles his teleported state with the state in the fifth node of the cluster state. As a result of this entanglement, the quadratures of Alice's and Bob's input modes, as well as the quadratures in the first and fifth modes of the cluster state, are transformed as follows:
\begin{multline}
    \hat{a}'_{in}=\hat{X}_{a_{in}'}+i\hat{Y}_{a_{in}'}\\
    \equiv \frac{\hat{x}_{a}+\hat{x}_{r,1}-g_{12} \hat{y}_{r,2}-g_{13} \hat{y}_{r,3}-g_{14} \hat{y}_{r,4}-g_{15}
   \hat{y}_{r,5}}{\sqrt{2}}+i\left(\frac{\hat{y}_{a}+\hat{y}_{r,1}+{g_{12} \hat{x}_{r,2}+g_{13} \hat{x}_{r,3}+g_{14} \hat{x}_{r,4}+g_{15}
   \hat{x}_{r,5}}}{\sqrt{2}}\right),
\end{multline}
\begin{multline}
    \hat{A}'_{1}=\hat{X}_{A_1'}+i\hat{Y}_{A_1'}\\
    \equiv \frac{\hat{x}_{a}-\hat{x}_{r,1}+g_{12} \hat{y}_{r,2}+g_{13} \hat{y}_{r,3}+g_{14} \hat{y}_{r,4}+g_{15}
   \hat{y}_{r,5}}{\sqrt{2}}+i\left(\frac{\hat{y}_{a}-\hat{y}_{r,1}-g_{12} \hat{x}_{r,2}-g_{13} \hat{x}_{r,3}-g_{14} \hat{x}_{r,4}-g_{15}
   \hat{x}_{r,5}}{\sqrt{2}}\right),
\end{multline}
\begin{multline}
  \hat{b}'_{in}= \hat{X}_{b_{in}'}+i\hat{Y}_{b_{in}'} \\
  \equiv \frac{\hat{x}_{b}+\hat{x}_{r,5}-g_{15} \hat{y}_{r,1}-g_{25} \hat{y}_{r,2}-g_{35} \hat{y}_{r,3}-g_{45}
   \hat{y}_{r,4}}{\sqrt{2}}+i\left(\frac{\hat{y}_{b}+\hat{y}_{r,5}+g_{15} \hat{x}_{r,1}+g_{25} \hat{x}_{r,2}+g_{35} \hat{x}_{r,3}+g_{45}
   \hat{x}_{r,4}}{\sqrt{2}}\right),
\end{multline}
\begin{multline}
  \hat{B}'_{2}= \hat{X}_{B_{2}'}+i\hat{Y}_{B_{2}'} \\
  \equiv \frac{\hat{x}_{b}-\hat{x}_{r,5}+g_{15} \hat{y}_{r,1}+g_{25} \hat{y}_{r,2}+g_{35} \hat{y}_{r,3}+g_{45}
   \hat{y}_{r,4}}{\sqrt{2}}+i\left(\frac{\hat{y}_{b}-\hat{y}_{r,5}-g_{15} \hat{x}_{r,1}-g_{25} \hat{x}_{r,2}-g_{35} \hat{x}_{r,3}-g_{45}
   \hat{x}_{r,4}}{\sqrt{2}}\right).
\end{multline}
The next step in the teleportation procedure is measurement. Alice and Bob measure the fields coming out of the beam splitter, and Charlie measures the field he has. The measurements in the scheme use homodyne detectors, which measure superpositions of quadratures of a field. The quantum amplitude of the photocurrent obtained from a homodyne measurement of a field with quadratures $\hat{x}$ and $\hat{y}$ can be written as follows:
 \begin{align} 
 \hat{i}=\beta _0 \left(\cos \theta \hat{x}+\sin \theta \hat{y}\right),
 \end{align}
where $\beta _0$ is the amplitude of the local oscillator, and ${\theta}$ is its phase. The quantum photocurrents obtained as a result of all measurements can be written as the following system of equations:
\begin{equation} \label{eq_cases}
    \begin{cases}
        \sin \theta_1 \hat{Y}_{a_{in}'}+\cos \theta_1 \hat{X}_{a_{in}'}=\frac{\hat{i}_{a_{in}}}{\beta_0},\\
   \sin \theta_2 \hat{Y}_{A_1'}+\cos \theta_2 \hat{X}_{A_1'}=\frac{\hat{i}_{A_1}}{\beta_0},\\
   \sin \theta_3 \hat{Y}_{b_{in}'}+\cos \theta_3 \hat{X}_{b_{in}'}=\frac{\hat{i}_{b_{in}}}{\beta_0},\\
   \sin \theta_4 \hat{Y}_{B_{2}'} +\cos \theta_4 \hat{X}_{B_{2}'}=\frac{\hat{i}_{B_2}}{\beta_0},\\
   \sin \theta_5 \hat{Y}_4+\cos \theta_5 \hat{X}_4=\frac{\hat{i}_{C}}{\beta_0}.
    \end{cases}
\end{equation}
Due to the presence of entanglements between the modes of the system under study, the measurement of some modes will affect others. To track this influence, it is necessary to solve the system of equations (\ref{eq_cases}) with respect to the quadratures $\hat{x}_{r,i}$ and substitute this solution into the expressions for the unmeasured quadratures $\lbrace \hat{X}_2,\hat{X}_3, \hat{Y}_2, \hat{Y}_3\rbrace$. As a result, these quadratures will be related to the quadratures of the input states of Alice and Bob. By analyzing these relations, it is easy to obtain restrictions on the weight coefficients and phases of the homodyne detectors that lead to the implementation of BQT. These restrictions are as follows:
\begin{align}
    &g_{15}=g_{25}=g_{13}=g_{23}=0, \label{14}\\
    &\theta_2=\theta_4=-\theta_1=-\theta_3=\frac{\pi}{4},\\
    &g_{12}=g_{35}=1, \label{16}\\
    &\theta_5=0.
\end{align}
Given (\ref{14}) and (\ref{16}), Fig. \ref{fig:TWT} presents the maximally general graph of the five-node cluster state suitable for implementing BQT in continuous variables.

Taking into account the obtained restrictions on the cluster state graph and also on the phases of the homodyne detectors, the relation between the output (unmeasured) quadratures and the input quadratures can be written in vector form as follows:
\begin{align} \label{out_5}
    \begin{pmatrix}
        \hat{X}_{out,B}\\
        \hat{X}_{out,A}\\
        \hat{Y}_{out,B}\\
        \hat{Y}_{out,A}
    \end{pmatrix}\equiv \begin{pmatrix}
        \hat{X}_{2}\\
        \hat{X}_3\\
        \hat{Y}_2\\
        \hat{Y}_3
    \end{pmatrix}=\begin{pmatrix}
        \hat{x}_{a}\\
        \hat{x}_b\\
        \hat{y}_a\\
        \hat{y}_b
    \end{pmatrix}+\frac{1}{\beta_0}\begin{pmatrix}
        -\hat{i}_{a_{in}}-\sqrt{2}g_{14}\hat{i}_C-\hat{i}_{A_1}\\
        -\sqrt{2} g_{45} \hat{i}_C-\hat{i}_{B_2}-\hat{i}_{b_{in}}\\
        \sqrt{2} g_{24} \hat{i}_C-\hat{i}_{A_1}+\hat{i}_{a_{in}}\\
        \sqrt{2} g_{34} \hat{i}_C-\hat{i}_{B_2}+\hat{i}_{b_{in}}
    \end{pmatrix}+\hat{\vec{e}},
\end{align}
where, to shorten the notation, a vector was introduced:
\begin{align} \label{error_vec}
   \hat{\vec{e}}= \begin{pmatrix}
        -2-g_{14}^2 && -g_{14}g_{24} && -g_{14}g_{34} && -g_{24} && -g_{14}g_{45}\\
        -g_{14}g_{45} && -g_{24}g_{45} && -g_{34}g_{45} && -g_{34} && -2-g_{45}^2\\
        g_{14}g_{24} && 2+g_{24}^2 && g_{24}g_{34} && g_{14} && g_{24}g_{45}\\
        g_{14}g_{34} && g_{24}g_{34} && 2+g_{34}^2 && g_{45} && g_{34}g_{45}
    \end{pmatrix}\begin{pmatrix}
        \hat{y}_{r,1}\\
        \hat{y}_{r,2}\\
        \hat{y}_{r,3}\\
        \hat{y}_{r,4}\\
        \hat{y}_{r,5}
    \end{pmatrix}.
\end{align}
In Eq. (\ref{out_5}), the first term on the right-hand side indicates that BQT has occurred. Alice has received Bob's state, and Bob has received Alice's state. The second term on the right-hand side of the Eq. (\ref{out_5}) is responsible for the measured photocurrents. After all the measurements have been taken, the photocurrent operators are replaced by real numbers corresponding to the measurement results. Since we know the photocurrent values in the experiment as well as the value of the cluster weight coefficients, we can compensate for all the classical quantities (real terms) in the output quadratures using the quadrature displacement operation. As a result of the displacement, we will have a vector of output quadratures consisting only of quantum operators and independent of the measurement results.
\begin{align} 
    \begin{pmatrix}
        \hat{X}_{out,B}\\
        \hat{X}_{out,A}\\
        \hat{Y}_{out,B}\\
        \hat{Y}_{out,A}
    \end{pmatrix}\equiv \begin{pmatrix}
        \hat{X}_{2}\\
        \hat{X}_3\\
        \hat{Y}_2\\
        \hat{Y}_3
    \end{pmatrix}=\begin{pmatrix}
        \hat{x}_{a}\\
        \hat{x}_b\\
        \hat{y}_a\\
        \hat{y}_b
    \end{pmatrix}+\begin{pmatrix}
        -2-g_{14}^2 && -g_{14}g_{24} && -g_{14}g_{34} && -g_{24} && -g_{14}g_{45}\\
        -g_{14}g_{45} && -g_{24}g_{45} && -g_{34}g_{45} && -g_{34} && -2-g_{45}^2\\
        g_{14}g_{24} && 2+g_{24}^2 && g_{24}g_{34} && g_{14} && g_{24}g_{45}\\
        g_{14}g_{34} && g_{24}g_{34} && 2+g_{34}^2 && g_{45} && g_{34}g_{45}
    \end{pmatrix}\begin{pmatrix}
        \hat{y}_{r,1}\\
        \hat{y}_{r,2}\\
        \hat{y}_{r,3}\\
        \hat{y}_{r,4}\\
        \hat{y}_{r,5}
    \end{pmatrix}.
\end{align}
We see that in the vector of output quadratures, in addition to the term responsible for the useful BQT, there is also a term depending on $\hat{y}_{r,j}$. As noted earlier, the operator $\hat{y}_{r,j}$ is a combination of squeezed $\hat{y}$-quadratures of the quantum oscillators used in the protocol. This term is responsible for the teleportation error associated with the use of real physical systems with a finite degree of squeezing. The greater the squeezing degree of the quantum oscillators used, the smaller the error value will be.

The presence of a small error in the results of teleportation is a feature of the scheme in continuous variables. In the protocol with continuous variables, we obtain deterministic teleportation with a small error. If we use the protocol in discrete variables, we get teleportation in a probabilistic manner. Thus, small errors are the price we pay for the determinism of the protocol.

\subsection{Bidirectional teleportation protocol errors on five-node cluster state in continuous variables}
In the previous section, we obtained that it is possible to implement  BQT on a five-node cluster state in continuous variables. In this case, teleportation occurs with an error that depends on the squeezed quadratures of the oscillators used to implement the protocol.

To talk about the reliability of the BQT protocol in continuous variables, we should minimize the teleportation errors. There are two ways to do this. The first way is to use oscillators with a large squeezing degree. However, it is challenging to create oscillators with a large squeezing degree experimentally. Currently, the maximum squeezing degree that can be achieved experimentally is 15.5 dB. For this reason, the degree of squeezing is a limited resource.

Another way to reduce the error is to optimize the implementation scheme of the \cite{Korolev_2020,Zinatullin_2022} protocol. The main goal of this approach is to reduce the contribution of the used quantum oscillators to the error. To minimize the errors, we analyze the error vector (\ref{error_vec}).  To minimize the error contribution from the squeezed oscillators, we need the matrix in front of the vector consisting of $\hat{y}_{r,j}$ quadratures to be maximally sparse. That is, it should contain the maximum number of zeros in it. Analyzing the error vector, we can distinguish three possible solutions: first $g_{14}=g_{45}=0$, $g_{24}\neq 0$, $g_{34}\neq 0$; second $g_{24}=g_{34}=0$, $g_{14}\neq 0$, $g_{45}\neq 0$; third $g_{24}=g_{34}=g_{14}=g_{45}=0$. Let us analyze each of them.

\begin{figure}[H]
    \centering
    \includegraphics{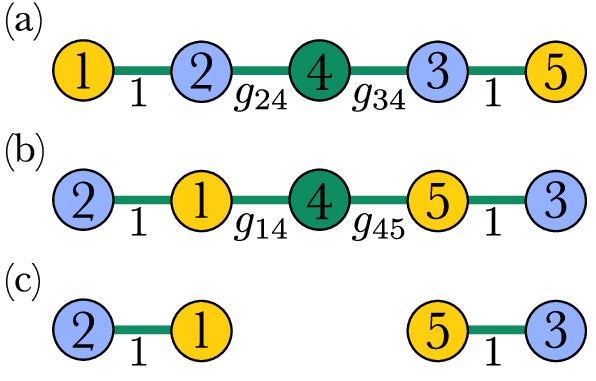}
    \caption{Three possible configurations of cluster states can be used to implement a bidirectional teleportation protocol with the least error. In the figure, nodes 1 and 5 (nodes marked in yellow) are the nodes to which the input states will be entangled, and nodes 2 and 3 (nodes marked in blue) are the nodes that will be in the output states.}
    \label{fig:min_error}
\end{figure}

The first solution ($g_{14}=g_{45}=0$, $g_{24}\neq 0$, $g_{34}\neq 0$) corresponds to the case of using a linear five-node cluster state, whose graph is shown in Fig. \ref{fig:min_error} (a).
\noindent The vector of error variances of  the BQT protocol realized on such a cluster state has the following form: 
\begin{align}
    \langle \delta \hat{\vec{e}}^2_1 \rangle=\begin{pmatrix}
        2\\
        2\\
        2+g_{24}^2\\
        2+g_{34}^2
    \end{pmatrix} \langle \delta \hat{y}_s^2 \rangle,
\end{align}
where the relation $\hat{\vec{y}}_r=\left(I+A_5^2\right)\hat{\vec{y}}_s$ was used, as well as the assumption that all used squeezed oscillators are independent and have the same squeezing degree (i.e. $\langle \delta \hat{y}_{s,i} \delta \hat{y}_{s,j} \rangle=\delta_{ij}\langle \delta \hat{y}_s \rangle$).

The second solution ($g_{24}=g_{34}=0$, $g_{14}\neq 0$, $g_{45}\neq 0$) also corresponds to the case of using the linear cluster state. The graph of this cluster is shown in Fig. \ref{fig:min_error} (b).  The difference between this case and the previous one is that in this configuration, the input states are entangled to the central nodes of the cluster. The vector of error variances for this case has the following form: 
\begin{align}
    \langle \delta \hat{\vec{e}}^2_2 \rangle=\begin{pmatrix}
        2+g_{14}^2\\
        2+g_{45}^2\\
        2\\
        2
    \end{pmatrix} \langle \delta \hat{y}_s^2 \rangle.
\end{align}
Comparing the first two cases, we see that, depending on the nodes to which the input states are entangled, the error will be larger in either $\hat{x}$ or $\hat{y}$-squares in the output states of Alice and Bob. Otherwise, these two cases are equivalent.

Finally, let us consider the third solution ($g_{24}=g_{34}=g_{14}=g_{45}=0$). The graph of such a cluster state is shown in Fig. \ref{fig:min_error} (c). We see that, unlike the previous two cases, this graph is not represented by a connected five-node graph, but by two two-node graphs distributed between Alice and Bob. This case can no longer be called BQT: it corresponds to two traditional teleportations. At the same time, the teleportation error in this configuration is the smallest and is given by the vector of error variances of the form:
\begin{align}
    \langle \delta \hat{\vec{e}}^2_3 \rangle=\begin{pmatrix}
        2\\
        2\\
        2\\
        2
    \end{pmatrix} \langle \delta \hat{y}_s^2 \rangle.
\end{align}
The obtained result has a simple explanation. The fewer nodes we use in the cluster state, the fewer sources of errors \cite{Korolev_2019}. Further reduction of nodes in the cluster will not yield results because, as we can see from the solutions we obtained, there is no three-node cluster state with which we can realize BQT in continuous variables. Thus, we can conclude that using two two-node cluster states realizes the BQT in continuous variables with the lowest error.

\section{Conclusion}
In this paper, we proposed a BQT protocol in continuous variables. We used a five-node cluster state as the main resource for BQT. Starting from a cluster state with a complete weighted graph, we showed that to realize bidirectional teleportation, some weight coefficients in the cluster state must be zero, some can be arbitrarily chosen, and some must be fixed. In other words, we have shown that it is possible to realize BQT in continuous variables, while there is arbitrariness in the choice of resource cluster state, namely arbitrariness in the choice of cluster state weight coefficients.

In addition, we have demonstrated that BQT in continuous variables occurs with errors. The variances of these errors are proportional to the variances of the squeezed quadratures of the quantum oscillators used, as well as to the weight coefficients of the cluster state graph. This means that by selecting the weight coefficients of the cluster state graph, we can reduce the teleportation error for a fixed squeezing of the used states. We have shown that the linear five-node cluster state allows realizing BQT with the lowest error among all five-node cluster states.

We have shown that the teleportation error will be even smaller if we use two two-node cluster states. In other words, employing two traditional teleportation protocols for BQT can minimize the error as much as possible. This result is easy to understand from a physical point of view. When implementing teleportation, each squeezed quantum oscillator serves as an error source, and minimizing the error is feasible with the least number of oscillators. Considering that a two-node cluster state is necessary for traditional teleportation, the minimum error of BQT is achieved using two two-node cluster states.

Thus, we have found that the best strategy for implementing BQT in continuous variables is to use two traditional teleportations. Moreover, such a protocol is simple from an experimental point of view, since it does not require the creation of complex multipartite states.

\vspace{0.5 cm}

This research was supported by the Theoretical Physics and Mathematics Advancement Foundation "BASIS" (Grants No. 24-1-3-14-1). SBK acknowledge support by the Ministry of Science and Higher Education of the Russian Federation on the basis of the FSAEIHE SUSU (NRU) (Agreement No. 075-15- 2022-1116).

\bibliography{bibliography}

\begin{thebibliography}{18}
\expandafter\ifx\csname natexlab\endcsname\relax\def\natexlab#1{#1}\fi
\expandafter\ifx\csname bibnamefont\endcsname\relax
  \def\bibnamefont#1{#1}\fi
\expandafter\ifx\csname bibfnamefont\endcsname\relax
  \def\bibfnamefont#1{#1}\fi
\expandafter\ifx\csname citenamefont\endcsname\relax
  \def\citenamefont#1{#1}\fi
\expandafter\ifx\csname url\endcsname\relax
  \def\url#1{\texttt{#1}}\fi
\expandafter\ifx\csname urlprefix\endcsname\relax\def\urlprefix{URL }\fi
\providecommand{\bibinfo}[2]{#2}
\providecommand{\eprint}[2][]{\url{#2}}

\bibitem[{\citenamefont{Bennett et~al.}(1993)\citenamefont{Bennett, Brassard,
  Cr\'epeau, Jozsa, Peres, and Wootters}}]{Bennett}
\bibinfo{author}{\bibfnamefont{C.~H.} \bibnamefont{Bennett}},
  \bibinfo{author}{\bibfnamefont{G.}~\bibnamefont{Brassard}},
  \bibinfo{author}{\bibfnamefont{C.}~\bibnamefont{Cr\'epeau}},
  \bibinfo{author}{\bibfnamefont{R.}~\bibnamefont{Jozsa}},
  \bibinfo{author}{\bibfnamefont{A.}~\bibnamefont{Peres}}, \bibnamefont{and}
  \bibinfo{author}{\bibfnamefont{W.~K.} \bibnamefont{Wootters}},
  \bibinfo{journal}{Phys. Rev. Lett.} \textbf{\bibinfo{volume}{70}},
  \bibinfo{pages}{1895} (\bibinfo{year}{1993}),
  \urlprefix\url{https://link.aps.org/doi/10.1103/PhysRevLett.70.1895}.

\bibitem[{\citenamefont{Zha et~al.}(2013)\citenamefont{Zha, Zou, Qi, and
  Song}}]{Zha_BDQT}
\bibinfo{author}{\bibfnamefont{X.-W.} \bibnamefont{Zha}},
  \bibinfo{author}{\bibfnamefont{Z.-C.} \bibnamefont{Zou}},
  \bibinfo{author}{\bibfnamefont{J.-X.} \bibnamefont{Qi}}, \bibnamefont{and}
  \bibinfo{author}{\bibfnamefont{H.-Y.} \bibnamefont{Song}},
  \bibinfo{journal}{International Journal of Theoretical Physics}
  \textbf{\bibinfo{volume}{52}}, \bibinfo{pages}{1740} (\bibinfo{year}{2013}),
  \urlprefix\url{https://doi.org/10.1007/s10773-012-1208-5}.

\bibitem[{\citenamefont{Fu et~al.}(2014)\citenamefont{Fu, Tian, and
  Hu}}]{Fu2014}
\bibinfo{author}{\bibfnamefont{H.-Z.} \bibnamefont{Fu}},
  \bibinfo{author}{\bibfnamefont{X.-L.} \bibnamefont{Tian}}, \bibnamefont{and}
  \bibinfo{author}{\bibfnamefont{Y.}~\bibnamefont{Hu}},
  \bibinfo{journal}{International Journal of Theoretical Physics}
  \textbf{\bibinfo{volume}{53}}, \bibinfo{pages}{1840} (\bibinfo{year}{2014}),
  \urlprefix\url{https://doi.org/10.1007/s10773-013-1985-5}.

\bibitem[{\citenamefont{Li and Nie}(2013)}]{Li_GHZ}
\bibinfo{author}{\bibfnamefont{Y.-h.} \bibnamefont{Li}} \bibnamefont{and}
  \bibinfo{author}{\bibfnamefont{L.-p.} \bibnamefont{Nie}},
  \bibinfo{journal}{International Journal of Theoretical Physics}
  \textbf{\bibinfo{volume}{52}}, \bibinfo{pages}{1630} (\bibinfo{year}{2013}),
  \urlprefix\url{https://doi.org/10.1007/s10773-013-1484-8}.

\bibitem[{\citenamefont{Yan}(2013)}]{Yan2013}
\bibinfo{author}{\bibfnamefont{A.}~\bibnamefont{Yan}},
  \bibinfo{journal}{International Journal of Theoretical Physics}
  \textbf{\bibinfo{volume}{52}}, \bibinfo{pages}{3870} (\bibinfo{year}{2013}),
  \urlprefix\url{https://doi.org/10.1007/s10773-013-1694-0}.

\bibitem[{\citenamefont{Duan et~al.}(2014)\citenamefont{Duan, Zha, Sun, and
  Xia}}]{Duan2014}
\bibinfo{author}{\bibfnamefont{Y.-J.} \bibnamefont{Duan}},
  \bibinfo{author}{\bibfnamefont{X.-W.} \bibnamefont{Zha}},
  \bibinfo{author}{\bibfnamefont{X.-M.} \bibnamefont{Sun}}, \bibnamefont{and}
  \bibinfo{author}{\bibfnamefont{J.-F.} \bibnamefont{Xia}},
  \bibinfo{journal}{International Journal of Theoretical Physics}
  \textbf{\bibinfo{volume}{53}}, \bibinfo{pages}{2697} (\bibinfo{year}{2014}),
  \urlprefix\url{https://doi.org/10.1007/s10773-014-2065-1}.

\bibitem[{\citenamefont{Chen}(2014)}]{Chen2014}
\bibinfo{author}{\bibfnamefont{Y.}~\bibnamefont{Chen}},
  \bibinfo{journal}{International Journal of Theoretical Physics}
  \textbf{\bibinfo{volume}{53}}, \bibinfo{pages}{1454} (\bibinfo{year}{2014}),
  \urlprefix\url{https://doi.org/10.1007/s10773-013-1943-2}.

\bibitem[{\citenamefont{Braunstein and van
  Loock}(2005)}]{Braunstein_RevModPhys.77.513}
\bibinfo{author}{\bibfnamefont{S.~L.} \bibnamefont{Braunstein}}
  \bibnamefont{and} \bibinfo{author}{\bibfnamefont{P.}~\bibnamefont{van
  Loock}}, \bibinfo{journal}{Rev. Mod. Phys.} \textbf{\bibinfo{volume}{77}},
  \bibinfo{pages}{513} (\bibinfo{year}{2005}),
  \urlprefix\url{https://link.aps.org/doi/10.1103/RevModPhys.77.513}.

\bibitem[{\citenamefont{Braunstein and Kimble}(1998)}]{Braunstein_telep_cv}
\bibinfo{author}{\bibfnamefont{S.~L.} \bibnamefont{Braunstein}}
  \bibnamefont{and} \bibinfo{author}{\bibfnamefont{H.~J.}
  \bibnamefont{Kimble}}, \bibinfo{journal}{Phys. Rev. Lett.}
  \textbf{\bibinfo{volume}{80}}, \bibinfo{pages}{869} (\bibinfo{year}{1998}),
  \urlprefix\url{https://link.aps.org/doi/10.1103/PhysRevLett.80.869}.

\bibitem[{\citenamefont{Korolev
  et~al.}(2020{\natexlab{a}})\citenamefont{Korolev, Golubeva, and
  Golubev}}]{Korolev_2020_1}
\bibinfo{author}{\bibfnamefont{S.~B.} \bibnamefont{Korolev}},
  \bibinfo{author}{\bibfnamefont{T.~Y.} \bibnamefont{Golubeva}},
  \bibnamefont{and} \bibinfo{author}{\bibfnamefont{Y.~M.}
  \bibnamefont{Golubev}}, \bibinfo{journal}{Laser Physics Letters}
  \textbf{\bibinfo{volume}{17}}, \bibinfo{pages}{035207}
  (\bibinfo{year}{2020}{\natexlab{a}}),
  \urlprefix\url{https://dx.doi.org/10.1088/1612-202X/ab6ffe}.

\bibitem[{\citenamefont{Vahlbruch et~al.}(2016)\citenamefont{Vahlbruch, Mehmet,
  Danzmann, and Schnabel}}]{Vahlbruch_2016}
\bibinfo{author}{\bibfnamefont{H.}~\bibnamefont{Vahlbruch}},
  \bibinfo{author}{\bibfnamefont{M.}~\bibnamefont{Mehmet}},
  \bibinfo{author}{\bibfnamefont{K.}~\bibnamefont{Danzmann}}, \bibnamefont{and}
  \bibinfo{author}{\bibfnamefont{R.}~\bibnamefont{Schnabel}},
  \bibinfo{journal}{Phys. Rev. Lett.} \textbf{\bibinfo{volume}{117}},
  \bibinfo{pages}{110801} (\bibinfo{year}{2016}),
  \urlprefix\url{https://link.aps.org/doi/10.1103/PhysRevLett.117.110801}.

\bibitem[{\citenamefont{Korolev
  et~al.}(2020{\natexlab{b}})\citenamefont{Korolev, Golubeva, and
  Golubev}}]{Korolev_2020}
\bibinfo{author}{\bibfnamefont{S.~B.} \bibnamefont{Korolev}},
  \bibinfo{author}{\bibfnamefont{T.~Y.} \bibnamefont{Golubeva}},
  \bibnamefont{and} \bibinfo{author}{\bibfnamefont{Y.~M.}
  \bibnamefont{Golubev}}, \bibinfo{journal}{Laser Physics Letters}
  \textbf{\bibinfo{volume}{17}}, \bibinfo{pages}{055205}
  (\bibinfo{year}{2020}{\natexlab{b}}),
  \urlprefix\url{https://dx.doi.org/10.1088/1612-202X/ab83ff}.

\bibitem[{\citenamefont{Korolev et~al.}(2018)\citenamefont{Korolev, Manukhova,
  Tikhonov, Golubeva, and Golubev}}]{Korolev_2018}
\bibinfo{author}{\bibfnamefont{S.~B.} \bibnamefont{Korolev}},
  \bibinfo{author}{\bibfnamefont{A.~D.} \bibnamefont{Manukhova}},
  \bibinfo{author}{\bibfnamefont{K.~S.} \bibnamefont{Tikhonov}},
  \bibinfo{author}{\bibfnamefont{T.~Y.} \bibnamefont{Golubeva}},
  \bibnamefont{and} \bibinfo{author}{\bibfnamefont{Y.~M.}
  \bibnamefont{Golubev}}, \bibinfo{journal}{Laser Physics Letters}
  \textbf{\bibinfo{volume}{15}}, \bibinfo{pages}{075203}
  (\bibinfo{year}{2018}),
  \urlprefix\url{https://dx.doi.org/10.1088/1612-202X/aac03e}.

\bibitem[{\citenamefont{Zinatullin et~al.}(2022)\citenamefont{Zinatullin,
  Korolev, Manukhova, and Golubeva}}]{Zinatullin_2022}
\bibinfo{author}{\bibfnamefont{E.~R.} \bibnamefont{Zinatullin}},
  \bibinfo{author}{\bibfnamefont{S.~B.} \bibnamefont{Korolev}},
  \bibinfo{author}{\bibfnamefont{A.~D.} \bibnamefont{Manukhova}},
  \bibnamefont{and} \bibinfo{author}{\bibfnamefont{T.~Y.}
  \bibnamefont{Golubeva}}, \bibinfo{journal}{Phys. Rev. A}
  \textbf{\bibinfo{volume}{106}}, \bibinfo{pages}{032414}
  (\bibinfo{year}{2022}),
  \urlprefix\url{https://link.aps.org/doi/10.1103/PhysRevA.106.032414}.

\bibitem[{\citenamefont{Gu et~al.}(2009)\citenamefont{Gu, Weedbrook, Menicucci,
  Ralph, and van Loock}}]{Gu}
\bibinfo{author}{\bibfnamefont{M.}~\bibnamefont{Gu}},
  \bibinfo{author}{\bibfnamefont{C.}~\bibnamefont{Weedbrook}},
  \bibinfo{author}{\bibfnamefont{N.~C.} \bibnamefont{Menicucci}},
  \bibinfo{author}{\bibfnamefont{T.~C.} \bibnamefont{Ralph}}, \bibnamefont{and}
  \bibinfo{author}{\bibfnamefont{P.}~\bibnamefont{van Loock}},
  \bibinfo{journal}{Phys. Rev. A} \textbf{\bibinfo{volume}{79}},
  \bibinfo{pages}{062318} (\bibinfo{year}{2009}),
  \urlprefix\url{https://link.aps.org/doi/10.1103/PhysRevA.79.062318}.

\bibitem[{\citenamefont{Bogoljubov}(1958)}]{Bogoljubov}
\bibinfo{author}{\bibfnamefont{N.~N.} \bibnamefont{Bogoljubov}},
  \bibinfo{journal}{Il Nuovo Cimento} \textbf{\bibinfo{volume}{7}},
  \bibinfo{pages}{794} (\bibinfo{year}{1958}).

\bibitem[{\citenamefont{Ukai et~al.}(2010)\citenamefont{Ukai, Yoshikawa, Iwata,
  van Loock, and Furusawa}}]{Ukai_1}
\bibinfo{author}{\bibfnamefont{R.}~\bibnamefont{Ukai}},
  \bibinfo{author}{\bibfnamefont{J.}~\bibnamefont{Yoshikawa}},
  \bibinfo{author}{\bibfnamefont{N.}~\bibnamefont{Iwata}},
  \bibinfo{author}{\bibfnamefont{P.}~\bibnamefont{van Loock}},
  \bibnamefont{and} \bibinfo{author}{\bibfnamefont{A.}~\bibnamefont{Furusawa}},
  \bibinfo{journal}{Phys. Rev. A} \textbf{\bibinfo{volume}{81}},
  \bibinfo{pages}{032315} (\bibinfo{year}{2010}).

\bibitem[{\citenamefont{Korolev et~al.}(2019)\citenamefont{Korolev,
  Dobrotvorskaia, Golubeva, and Golubev}}]{Korolev_2019}
\bibinfo{author}{\bibfnamefont{S.~B.} \bibnamefont{Korolev}},
  \bibinfo{author}{\bibfnamefont{A.~N.} \bibnamefont{Dobrotvorskaia}},
  \bibinfo{author}{\bibfnamefont{T.~Y.} \bibnamefont{Golubeva}},
  \bibnamefont{and} \bibinfo{author}{\bibfnamefont{Y.~M.}
  \bibnamefont{Golubev}}, \bibinfo{journal}{Laser Physics Letters}
  \textbf{\bibinfo{volume}{16}}, \bibinfo{pages}{075204}
  (\bibinfo{year}{2019}),
  \urlprefix\url{https://dx.doi.org/10.1088/1612-202X/ab189a}.

\end{thebibliography}

\end{document}